\documentclass[a4paper]{article}

\usepackage[left = 3.5 cm, right = 3.5 cm]{geometry}
\usepackage[T1]{fontenc} %% Use modern font encodings
\usepackage[utf8]{inputenc}
\usepackage{lmodern}
\usepackage{bm} %% for bold vectors
\usepackage{amsmath, amsfonts, amssymb}
\usepackage{graphicx}
\usepackage{hyperref}
\usepackage{upgreek}
\usepackage[super, sort&compress]{natbib}
\usepackage{authblk}
\usepackage{todonotes, soul, xcolor}
\usepackage{lineno}
\usepackage{setspace}

\usepackage{csquotes}
\usepackage{comment}

\newcommand{\stk}[1]{\ensuremath{\left(#1\right)}}

\author[1, 3, 5]{Konr\'{a}d Kandrai}
\author[2]{Andr\'{a}s Balogh}
\author[1, 5]{D\'{o}ra Varga}
\author[1]{M\'{a}rton Szendr\H{o}}
\author[1]{Andr\'{a}s P\'{a}link\'{a}s}
\author[2]{J\'{o}zsef Cserti}
\author[2]{J\'{a}nos Koltai}
\author[1, 4, 5]{Kriszti\'{a}n M\'{a}rity}
\author[1]{P\'{e}ter Vancs\'{o}}
\author[1]{G\'{e}za I. M\'{a}rk}
\author[6]{Alexandre Mayer}
\author[1, 2, 5]{Zolt\'{a}n Tajkov}
\author[1, 3]{Levente Tapaszt\'{o}}
\author[1, 5 *]{P\'{e}ter Nemes-Incze}

\affil[1]{Hungarian Research Network, HUN-REN Centre for Energy Research, Institute of Technical Physics and Materials Science, 1121 Budapest, Hungary}
\affil[2]{ELTE, Eötvös Loránd University, 1117 Budapest, Hungary}
\affil[3]{Department of Physics, Institute of Physics, Budapest University of Technology and Economics, M\H{u}egyetem rkp. 3., H-1111 Budapest, Hungary}
\affil[4]{\'{O}buda University, 1034 Budapest, Hungary}
\affil[5]{MTA - HUN-REN EK Lendület ``Momentum'' Topology in Nanomaterials Research Group, 1121 Budapest, Hungary}
\affil[6]{Department of Physics, Namur Institute for Complex Systems (naXys), University of Namur, Namur, Belgium}
\affil[*]{\small \emph{corresponding author, email: nemes.incze.peter@ek.hun-ren.hu}}

\title{One-way electron mirrors from single-layer stacking shifts in rhombohedral graphite}

\begin{document}

\maketitle

\doublespacing

\textbf{
One-way mirrors or asymmetric scatterers are key components of photonic systems, protecting sources from back-reflections and routing light along a chosen path.
Acoustic systems can also be engineered to show asymmetric reflection.
However, a condensed matter platform acting as a one-way mirror without breaking linearity, time-reversal symmetry or energy conservation has been lacking.
One route to achieving this is mode selection, where a probe coupled to only some modes sees a boundary as a mirror from one side and as transparent from the other.
Here we show that single-layer stacking shifts in six-layer rhombohedral graphite provide this platform, acting as one-way mirrors for electrons.
In scanning tunneling microscopy, 10 nm scale interference fringes appear only on the stacking fault side of a wall, tracking its Dirac band, the rhombohedral side remaining transparent.
Layer-resolved tight-binding and wave-packet dynamics calculations reproduce this and trace the asymmetry to the mismatched layer polarization across the wall, with the tip supplying the selective coupling.
Because stacking solitons glide under external fields, this mirror forms a movable element of electron optics defined by the crystal structure alone.
}

\section*{Introduction}

A one-way mirror for waves is an asymmetric scatterer, its response depends on the direction from which the wave arrives.
Devices built on this principle are basic elements of electronic, photonic, and acoustic technology: the diode rectifying an alternating current~\cite{Shockley1949}, the Faraday optical isolator~\cite{Aplet1964} protecting a laser from its own back-reflections~\cite{Jalas2013}, and the acoustic panels that absorb sound arriving from one side and reflect it from the other~\cite{Du2021-vb, Jimenez2017}.
Their directionality results from breaking one of three conditions: linearity (the diode), energy conservation (the absorbing panel), or time-reversal symmetry (the magneto-optical isolator)~\cite{Jalas2013, Maznev2013, Caloz2018}.
Mode selection provides an exception.
If several modes coexist at the same energy and the probe couples to only some of them, a boundary can act as a mirror from one side and appear transparent from the other.
Here we demonstrate scattering asymmetry with none of the above conditions broken: in the elastic scattering of quasiparticles at fixed energy, in a system with time-reversal symmetry.
Using scanning tunneling microscopy (STM) we show directional reflection asymmetry of electron waves at stacking boundaries of rhombohedral graphite (RG), a one-way mirror for electrons, via quasiparticle interference (QPI) measurements.
The measurements show interference fringes in the local density of states (LDOS) on one side of the wall only.
Layer-resolved tight-binding and wave packet dynamics calculations show that this one-sidedness is a difference in reflection and not only in the fringe visibility.
We establish that the source of the reflection asymmetry is mode selection, which allows asymmetric scattering without breaking any of the three conditions.
At a given energy, a quasiparticle can propagate in several modes, set apart by their polarization, their spatial shape, or, in a layered crystal, by the distribution of the wave-function weight over the layers.
A boundary can convert one mode into another of the same energy upon reflection, and a source or detector can couple to only some of these modes.
A wave arriving from one side may then be scattered into modes invisible to the probe (STM tip), in reflection or in transmission, while from the other side the reflection returns in a visible mode.
Scattering reciprocity~\cite{Buttiker1986-bg, Buttiker1988-lz} is respected throughout: it equates the two directions of every transmission between modes, and allows the reflections seen through a subset of the modes to differ.

So far, one-way mirrors that break none of the three conditions have been realized only in photonic and acoustic systems.
Photonic gratings route the two incidence directions into different diffraction orders~\cite{Serebryannikov2009, Foteinopoulou2022}, and an inverse-designed silicon waveguide passes light in one direction only by converting between its spatial modes, a reciprocal optical diode for integrated photonics~\cite{Callewaert2016}.
Acoustic gratings do the same for sound~\cite{Li2011, He2011}, and elastic metamaterials achieve circulator-like routing and unidirectional focusing of guided waves, for signal routing and ultrasonic sensing~\cite{Hu2025, Zhu2025}.
In all of these, both ingredients, the mode structure and the selective coupling, are engineered into the device by fabrication.
A condensed matter realization, in which the modes and the boundary that converts them are supplied by the crystal itself, has been missing.
RG electronic states with wave function weights on different layers, as well as differences in sublattice polarization, provide the modes, while the stacking boundary converts them~\cite{Mark2025-ek}.
The selective coupling is supplied by the STM tip, since it couples to the top graphene layer and so addresses only the modes with wave function weight there.

Stacking and twisting of graphene layers result in new emergent physics, especially in flat band systems such as magic angle twisted bilayer graphene~\cite{Cao2018-hm}.
RG needs no twisting: its band structure is tuned by the layer number~\cite{Ho2016-lu, Seifert2024-ei, Slizovskiy2019-sm} and the stacking configuration~\cite{Atri2024-sv, Wirth2022-wq, Garcia-Ruiz2023-zu}.
Its surface-localized flat band hosts correlated phases~\cite{Zhang2024-wn, Liu2025-kd}, from superconductivity~\cite{Kumar2026-la} to the fractional quantum anomalous Hall effect~\cite{Lu2024-ek}.
Stacking sequences of RG are separated by domain wall solitons~\cite{Koshino2013-bw, Vizner-Stern2024-fc, Mark2025-ek}, with a lateral size in the 10 nm range~\cite{Yin2017-go, Yin2023-as, Yankowitz2014-bf, Hattendorf2013-vo}.
Apart from allowing the relaxation of in-built mechanical deformation stemming from the change in stacking, these domain walls host topological valley channels in gated bilayer graphene~\cite{Martin2008-km, Ju2015-fq, Yin2016-yi}, can organize into networks~\cite{Rickhaus2018-as, Efimkin2018-uz, Song2026-fw} and can be manipulated~\cite{Jiang2018-ci}.
Because they are smooth on the atomic scale, they contribute to intravalley scattering~\cite{Mark2025-ek, Yin2023-as} without the intervalley component produced by atomically sharp defects such as grain boundaries~\cite{Nemes-Incze2013-ku}.
Interference fringes at such a domain wall have been observed by STM~\cite{Yin2023-as}, and electron transmission across stacking boundaries has been calculated~\cite{Nam2015}, but the direction dependence of the reflection has not been measured.

Here we present STM measurements on a six-layer graphite flake, where hexalayer RG domains form lateral junctions with stacking fault domains of the same six-layer thickness: pentalayer RG capped by a sixth graphene layer in the hexagonal position.
In this stacking sequence, the top graphene layer we measure via STM is in the hexagonal position, as opposed to the rhombohedral one (see Fig.~\ref{fig:Intro}a).
Using QPI measurements we show that the domain wall contributes to intravalley scattering, and that it acts as an asymmetric scatterer between the hexalayer RG and the stacking fault, reflecting the electrons of the Dirac band of the stacking fault.
Between two solitons these states form standing waves within the RG gap, in a Fabry--P\'erot geometry whose mirrors are single-layer shifts in stacking.
The result is an electron optics building block, based only on the crystallographic stacking sequence of RG, that is switchable via external fields~\cite{Yankowitz2014-bf, Yeo2025-qf, Vizner-Stern2024-fc}.

\section*{Results}

\begin{figure}[!ht]
    \centering
    \includegraphics[width=1\textwidth]{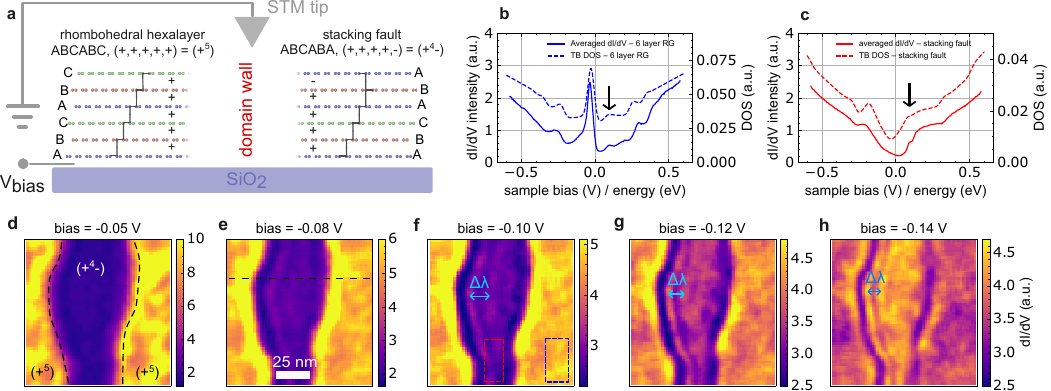}
	\caption{\textbf{Bias-dependent STM/STS mapping across a rhombohedral-stacking-fault domain wall.}
		\textbf{(a)} Schematic illustration of the STM experiment performed on hexalayer graphite containing a stacking domain wall.
		The stacking sequences are given in the H\"agg sign notation defined in the text: \stk{+^5} denotes the six-layer rhombohedral (ABCABC) stacking, while \stk{+^4 -} denotes the stacking fault (ABCABA), in which the sign of the topmost interface is reversed by a lateral shift of the top layer toward hexagonal (Bernal) stacking.
		The schematic serves to illustrate the local stacking configuration probed by STM.
		\textbf{(b, c)} Spatially averaged experimental dI/dV spectra (colored continuous curves) extracted from the regions marked in panel f, shown together with tight-binding calculations of the surface density of states (colored dashed curves).
		Bottom layer flat band DOS peak shown by black arrow.
		Calculated DOS is offset for clarity.
		\textbf{(d-h)} Spatially resolved differential conductance (dI/dV) maps acquired on the same area at different bias voltages, highlighting the evolution of electronic states across the stacking domain wall.
		The domain wall positions are indicated by black dashed lines on panel d.
		Panel e includes a scale bar of 25 nm.
		The black dashed line in panel e marks the location of the line spectrum presented in Fig.~\ref{fig:scattering1}a.
		\textbf{(f)} Red and blue rectangles indicate regions within the stacking fault and rhombohedral domains, respectively, from which dI/dV spectra were spatially averaged in panels b and c.
	}
    \label{fig:Intro}
\end{figure}

Figure~\ref{fig:Intro}a schematically shows the measurement setup, with the atomic structure of the hexalayer RG and the stacking fault region, both six graphene layers thick. 
As each graphene layer can occupy either the BA or AB position, with respect to the layer below, we introduce here a convenient shorthand notation for the graphite stacking sequence.
Here we do not label the graphene layers, but the more intrinsic property of the relative position of the layer to the one below, with AB being denoted by ``$+$'' and BA as ``$-$''.
Thus the hexalayer RG is \stk{+,+,+,+,+} and the stacking fault region, with the top graphene layer in the hexagonal position, is \stk{+,+,+,+,-}.
This sign sequence follows the H\"agg notation of polytypism~\cite{Guinier-Chairman-1984-zr}: it is read from the bottom layer up, so the last sign belongs to the topmost interface, the one closest to the STM tip.
Unlike the letter notation, where ABCABC and BCABCA name the same crystal, the sign sequence is unique up to an overall sign, which corresponds to a $180^{\circ}$ rotation of the crystal about the stacking axis.
In a compressed notation, runs of equal signs are abbreviated by exponents: the hexalayer RG reads $\stk{+^{5}}$ and the stacking fault region $\stk{+^{4}-}$.
The two sequences differ in a single sign, that of the topmost interface: the domain walls in Fig.~\ref{fig:Intro} are the lines along which this sign flips.

We identify the RG and stacking fault regions via electronic Raman scattering (ERS)~\cite{Palinkas2024-sm} (see Supplementary Section~S1) and by their unique surface LDOS fingerprint.
Figure~\ref{fig:Intro}b,c shows tunneling conductance measurements (dI/dV) of the two areas, with the RG displaying its characteristic flat-band surface state.
In the stacking fault region, this surface state is buried one layer deep in the structure; thus it is not visible in the dI/dV measurement.
Furthermore, smaller LDOS peaks in both structures, stemming from van Hove singularities at the gap edges, act as fingerprints of the stacking structure in STM~\cite{Hattendorf2013-vo, Palinkas2024-sm, Zhang2025-ut}.
Tight binding calculations of the surface DOS of each structure identify the stacking sequences as $\stk{+^{5}}$ and $\stk{+^{4}-}$.

Figure~\ref{fig:Intro}d-h shows dI/dV maps, at various sample voltages, of a stacking fault region bounded by two domain walls, marked by black dashed lines in Fig.~\ref{fig:Intro}d.
The maps exhibit 10 nm scale standing waves in the LDOS of the stacking fault region, between the two walls.
These ripples in the LDOS show a clear dependence on the sample bias at which they are measured, indicating intravalley scattering as their source.
The two domain walls thus act as the mirrors of a Fabry--P\'erot cavity for the surface states of the stacking fault, as step edges do for the surface state of metals, for example Ag(111)~\cite{Burgi1998-fg}.
The walls are curved and the width of the region varies from 40 to 70~nm, so the cavity has no single length and we do not resolve the discrete mode spectrum seen in resonators with parallel walls, such as terraces bounded by straight step edges on Ag(111)~\cite{Burgi1998-fg} and on rhombohedral trilayer graphene~\cite{Han2023}.

\begin{figure}[!ht]
    \centering
    \includegraphics[width=1\textwidth]{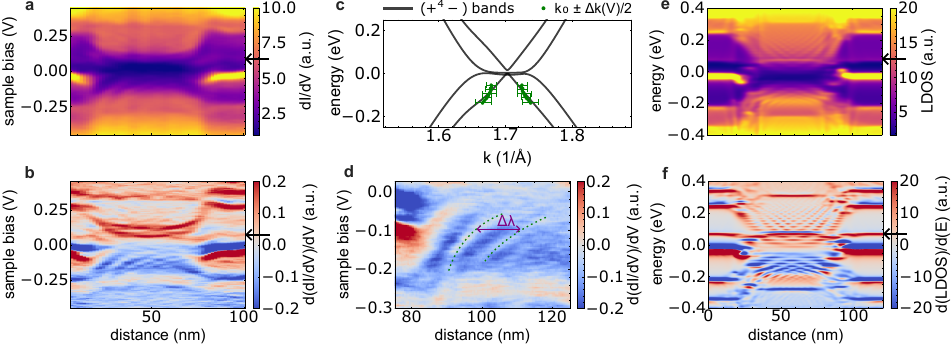}
	\caption{\textbf{Energy-dependent quasiparticle interference and extraction of scattering wave vectors.}
		\textbf{(a)} Differential conductance (dI/dV) line spectra measured along the trajectory indicated by the black dashed line in Fig.~\ref{fig:Intro}e, crossing the stacking-fault region between the walls.
		\textbf{(b)} Numerical derivative of the dI/dV line spectra in panel a, taken along the sample bias axis.
		This representation enhances the visibility of interference patterns and quasiparticle scattering fringes.
		\textbf{(c)} Extraction of scattering wave vectors from dI/dV signal oscillations.
		The measured interference wavelengths are converted to scattering vectors $\Delta k = 2 \pi / \lambda$.
		The experimentally obtained $k_{0} \pm \Delta k/2$ values are overlaid on the calculated band structure of $\stk{+^4-}$, where the Dirac point $k$ vector is $k_{0}$.
		\textbf{(d)} The $\stk{+^4-}$ area, with a domain wall on the left side, from which the scattering wave vectors in c are extracted.
		The wavelength $\lambda$ is extracted as the distance between adjacent intensity maxima/minima and shows a clear sample bias dependence.
		\textbf{(e and f)} Map of the energy-dependent local density of states (LDOS) corresponding to panels a and b, calculated in our tight-binding model.
		The LDOS signal of the bottom flat band shows up in the doped RG, shown by black arrows in the panels.
	}
	 \label{fig:scattering1}
\end{figure}

Turning our attention to the LDOS ripples observed in the $\stk{+^{4}-}$ region, we plot the gradient of the dI/dV signal with respect to the sample bias in Fig.~\ref{fig:scattering1}b.
This representation visualizes the observed interference fringes with higher contrast.
We attribute these fringes to the backscattering and interference of surface electronic states in the surface modes of the $\stk{+^{4}-}$ structure.
On the surface of the stacking fault there is a single Dirac cone, localized on the top layer, whose two branches cross near the Fermi level (see Fig.~\ref{fig:scattering1}c).
Examining the interference fringes close to a single domain wall (Fig.~\ref{fig:scattering1}d), the sample bias dependence, and thus the energy dependence, of the wavelength ($\lambda$) becomes clear.
Converting the wavelengths to distances in reciprocal space ($\Delta k = 2\pi/\lambda$) and overlaying them on the band structure plot in Fig.~\ref{fig:scattering1}c, the wave vectors fall onto the Dirac dispersion of the $\stk{+^{4}-}$ surface mode.
No interference fringes are resolved on the $\stk{+^{5}}$ side of the line spectra.
For more examples see Supplementary Section~S2.

\begin{figure}[!ht]
    \centering
    \includegraphics[width=1\textwidth]{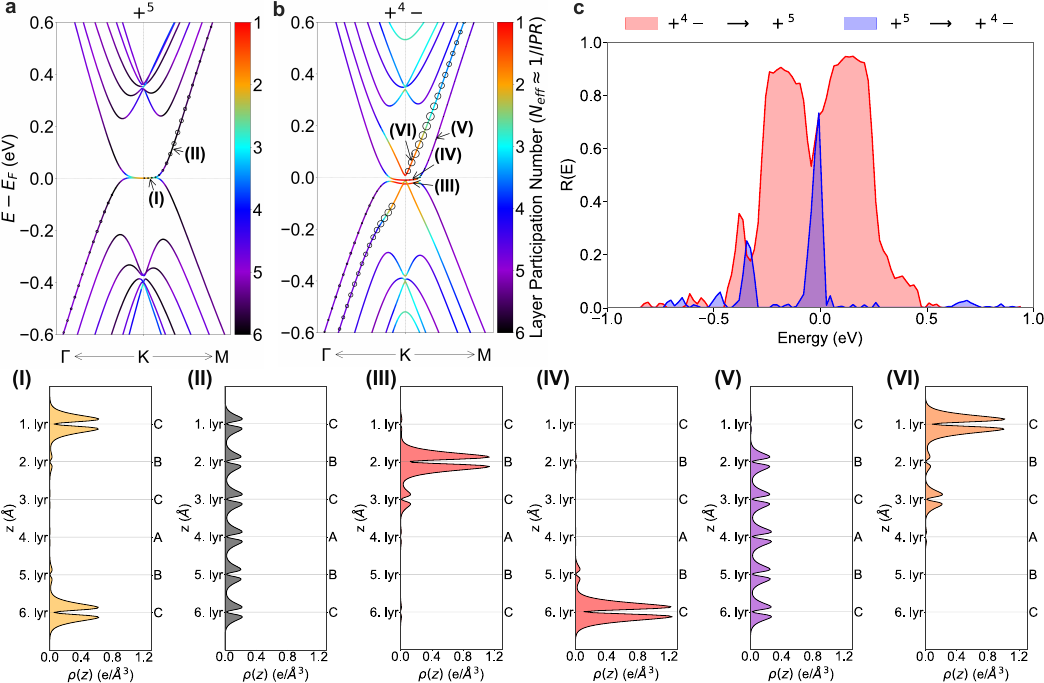}
	\caption{\textbf{Wave-packet dynamics simulations of layer polarization and directional scattering at rhombohedral-stacking-fault interfaces.}
		\textbf{(a)} Layer-resolved weight of the electronic states in rhombohedral \stk{+^5} stacking.
		\textbf{(b)} Corresponding layer polarization in the stacking-fault configuration \stk{+^4 -}.
		The localization of the electronic states onto one or multiple layers is displayed as a color scale.
		The band structures in a and b are obtained from density functional theory (DFT) calculations.
		Black circles on the bands illustrate, for visual purposes, the spectral weight of the states populated by the wave packet injected from the lead, with the radius of the circles proportional to the population~\cite{Mark2025-ek}.
		\textbf{(c)} Energy-resolved reflection probability of the populated modes incident on the domain wall from the two sides, calculated as $R(E) = 1 - I_{\mathrm{out}}(E)/I_{\mathrm{in}}(E)$, following the method of Ref.~\cite{Mark2025-ek}.
		$R(E)$ sums the reflection into all modes and layers.
		The spectra show a pronounced directional asymmetry: strong reflection for \stk{+^4 -}$\,\rightarrow\,$\stk{+^5} incidence and high transmission in the opposite direction.
		\textbf{(I--VI)} Layer-resolved charge density $\rho(z)$ of the states marked I--VI in a and b; the letters give the stacking position of each layer, layer 1 being the topmost.
		In the \stk{+^5} domain the flat-band state I is localized on the two outer layers, while the dispersive state II is spread over all six layers.
		In the \stk{+^4 -} domain the Dirac-cone state VI is localized predominantly on the top layer, state V has negligible weight on the top layer, and the flat-band state III is buried on layer 2, one layer below the surface, with state IV on the bottom layer.
		The asymmetry in c arises from the mismatch in layer polarization, as well as the selective population of transport channels, leading to mode-dependent coupling across the interface.
	}
	 \label{fig:wpd}
\end{figure}

Whether reflection at the domain wall produces observable interference depends on the band structure of the states involved.
A state backscattered at energy $E$ modulates the local density of states as $\cos(2k(E)x + \varphi)$, with a fringe period $\lambda = \pi/k(E)$ set by the wavevector of the band, corresponding to the scattering vector $\Delta k = 2k = 2\pi/\lambda$~\cite{Crommie1993-lo, Jeandupeux1999-tj}.
The dI/dV signal averages this modulation over an energy window of $\Delta E \approx 7$~meV, set by our lock-in modulation (3.75~mV) and the temperature ($3.5\,k_BT \approx 2.7$~meV at 9~K).
States within this window contribute wavevectors spread by $\delta k = \Delta E/\hbar v_g$, so the fringes dephase beyond a distance $x_c \approx \pi \hbar v_g/(2\Delta E)$~\cite{Jeandupeux1999-tj}.
Therefore the visibility of the fringes is set by the group velocity.
For the Dirac band of the stacking fault ($v_g = 5\times10^{5}$~m/s), the fringe period is $\pi\hbar v_g/|E - E_D| \approx 10$~nm at 100 mV sample bias (Fig.~\ref{fig:scattering1}d), and the coherence length is $x_c \approx 73$~nm, spanning the 40--70~nm wide stacking-fault regions in Fig.~\ref{fig:Intro}d-h.
Thus we expect to observe reflections at the domain wall in the dI/dV maps for the $\stk{+^{4}-}$ stacking fault area in the energy range of the $\stk{+^{5}}$ gap ($-0.22$ to $0.25$~eV).
For energies around the flat band of the RG we have the opposite limit.
It disperses by $W_\mathrm{FB} \approx 10$~meV over a momentum extent of $k_\mathrm{FB} \approx 0.06$~\AA$^{-1}$, corresponding to an average group velocity of $\bar{v}_\mathrm{FB} = W_\mathrm{FB}/\hbar k_\mathrm{FB} \approx 2.5\times10^{4}$~m/s, twenty times below that of the Dirac band.
The corresponding fringes have periods down to $\pi/k_\mathrm{FB} \approx 5$~nm and a coherence length of just $x_c \approx 4$~nm.
This coherence length is at or below the $\sim 10$~nm structural width of the soliton itself.
In the energy range of the flat band we therefore do not expect to see interference fringes.
This is apparent in the dI/dV line scans shown in Fig.~\ref{fig:scattering1} and in the measurements of a thin RG ribbon presented in Supplementary Section~S2.
The absence of fringes on the $\stk{+^{5}}$ side in the flat-band energy range could therefore be a visibility effect alone, owing to the difference in group velocities.
The suppression of interference fringes in experiments in the energy range of the dispersive part of the flat band (marked by II in Fig.~\ref{fig:wpd}a) requires a more detailed analysis using tight-binding calculations.

\begin{figure}[!ht]
    \centering
    \includegraphics[width=1\textwidth]{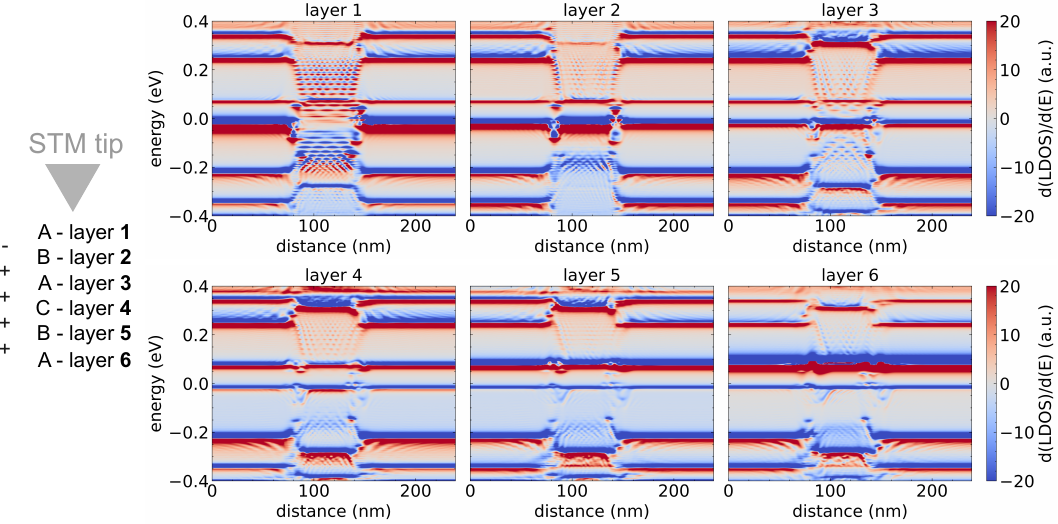}
	\caption{\textbf{Layer resolved LDOS at the domain wall.}
		Comparison of the calculated LDOS of each graphene layer across the six layer region.
		Inset on the left, shows the numbering of the layers, with layer 1 being measured by the STM tip.
		Interference fringes in the $\stk{+^4-}$ region are present most prominently in the top layer (layer 1), as also shown in Fig.~\ref{fig:scattering1}, and they are strongly suppressed in the $\stk{+^5}$ area, within the RG gap, $-0.22$ to $0.25$ eV.
	}
	\label{fig:layerLDOS}
\end{figure}

To treat the interference effects more generally and to reproduce the surface LDOS we measure, we turn to tight-binding (TB) calculations (see Methods).
We build the atomic structure of the whole sample, including the domain walls, in LAMMPS, and calculate the surface LDOS via a Slater-Koster tight-binding approach~\cite{Balogh2026-ni}.
These calculations capture the experimental surface LDOS well, see for example Fig.~\ref{fig:Intro}b,c.
To reproduce the small experimental spectroscopic features, we needed to consider the doping of the RG flake by the SiO$_2$ substrate.
As seen from the slight shift of the flat band from the Fermi level in the experimental spectra (Fig.~\ref{fig:Intro}b and Fig.~\ref{fig:scattering1}a), there is a slight doping of the sample, which can switch from n to p type over the surface~\cite{Zhang2009-gn, Hagymasi2022-hi} (see Supplementary Section~S4).
In these calculations the doping is not constant over all graphene layers of the RG, but we consider the decay of the electric field present at the substrate--RG interface within the sample, due to the non-negligible screening of the RG; for more details see Methods and Supplementary Section~S3.
The TB calculations not only reproduce the interference fringes on the $\stk{+^{4}-}$ surface but also the lack thereof in the defect free RG, see Fig.~\ref{fig:scattering1}e,f.
% The fringes remain strongest on the $\stk{+^{4}-}$ side for all domain wall orientations we have measured, between 12$^{\circ}$ and 24$^{\circ}$ from the zigzag direction, and in TB calculations at different wall orientations (Supplementary Section~S5).
One result of this slight surface doping is the electron--hole asymmetry in the observed interference fringes.
This asymmetry is due to the small wave function weight of the bottom flat band appearing on the top surface as a weak LDOS peak, marked by black arrows in Fig.~\ref{fig:Intro}b,c and Fig.~\ref{fig:scattering1}e,f.
This LDOS peak tends to obscure the interference fringes in the vicinity of the energy where it appears.
For more details see Supplementary Section~S4.
The lack of interference fringes on the $\stk{+^{5}}$ area is not just a result of the small group velocity near the flat band energies.
In the $\stk{+^{5}}$ area, the strongly dispersing branches of the flat band are delocalized over all graphene layers (see Fig.~\ref{fig:wpd}a, II).
Here, our group velocity argument fails, since the dispersion of these bands is similar to the surface Dirac band dispersion of the $\stk{+^{4}-}$ area.
In Fig.~\ref{fig:layerLDOS} we compare the LDOS of each graphene layer in the six-layer region, and observe that the interference fringes are absent also in the deeper layers of the defect free RG.
Thus, we need to look beyond the TB model we have presented thus far to uncover the origin of this asymmetric scattering.

The TB description can reproduce the equilibrium LDOS, but it does not give us insight into the mechanism of the observed reflection asymmetry.
We address this by using wave packet dynamics (WPD) calculations to inject a Gaussian wave packet either from the $\stk{+^{5}}$ or the $\stk{+^{4}-}$ side, with graphene leads connected to the top layer, which is probed by the STM.
The supplementary movies show a time series of the wave packet propagation from both sides of the domain wall, demonstrating qualitatively that the wave packet incident from the $\stk{+^{4}-}$ area suffers greater reflection than that incident from the $\stk{+^{5}}$ side (see Supplementary Section~S5).

To understand the scattering processes involved, in Fig.~\ref{fig:wpd}a,b we plot the density functional theory (DFT) band structures of the two sides of the domain wall.
Coloring of the bands shows the wave function localization on the number of graphene layers ($N_{\mathrm{eff}}$) in the structure, as represented by the participation ratio~\cite{Bell1970-pr, Edwards1972-pr}.
The color bar in the panels shows that the flat band of the $\stk{+^{5}}$ RG is localized to two graphene layers (top and bottom), while the Dirac cone of the $\stk{+^{4}-}$ is localized predominantly on one graphene layer, the top.
In the WPD calculations we simulate the STM tip by injecting the initial wave packet into the top graphene layer on both sides, via graphene leads (see Supplementary Section~S5 and Supplementary Fig.~S5).
Black circles in the bands show how much the graphene leads populate each band, revealing that the initial wave packet populates the $\stk{+^{5}}$ flat band and its $E \sim k^6$ dispersive part, as well as the Dirac band of the $\stk{+^{4}-}$.
Some layer resolved charge densities at selected energies are also plotted in panels I to VI.
These plots show a pronounced difference in the layer resolved charge density on the two sides.
In particular, in the dispersive part of the flat band, within the $\stk{+^{5}}$ gap, the wave function is delocalized over all layers (panel II), while in the case of the $\stk{+^{4}-}$ the charge density is localized predominantly on the top graphene layer (panel VI).
This mismatch is at the root of the asymmetric scattering observed at energies within the gap of the RG.
To quantify the direction dependence of the scattering, we calculate the energy-resolved reflection $R(E)$ at the domain wall, following the method of Ref.~\cite{Mark2025-ek}.
A wave packet is injected from a graphene lead into the topmost layer of the junction, and the probability current it carries is followed in time.
A time--energy Fourier transform of the wave function resolves the process in energy.
The reflection probability is obtained as $R(E) = 1 - I_{\mathrm{out}}(E)/I_{\mathrm{in}}(E)$, where $I_{\mathrm{out}}$ is the current transmitted through the full junction and $I_{\mathrm{in}}$ is the current carried by the same wave packet in a reference system without the domain wall, summed over all of its populated modes.
This normalization removes the mode filtering of the injecting contact: $R(E)$ measures the reflection caused by the domain wall alone.
The two incidence directions are compared on the same footing by mirroring the relaxed junction geometry, which interchanges the $\stk{+^{5}}$ and $\stk{+^{4}-}$ domains while preserving the atomic structure of the wall.
The resulting $R(E)$ (Fig.~\ref{fig:wpd}c) counts the reflected probability current in all modes and all layers of the junction.
The calculated $R(E)$ is not directly related to the measured interference fringes.
The fringe amplitude in dI/dV measures a projection of this reflection: only the part that returns in the same band with wave function amplitude present on the top layer produces an interference pattern visible to the tip.
On the $\stk{+^{4}-}$ side this distinction is mild, since the Dirac band is polarized on the top layer: the observed fringes capture the dominant part of the calculated reflection.
On the $\stk{+^{5}}$ side the distinction is stronger, since reflection of the modes delocalized over all layers is less pronounced at the tip.
However, our TB calculations resolve the LDOS in each graphene layer separately, and show interference fringes in none of the six layers of the $\stk{+^{5}}$ domain (Fig.~\ref{fig:layerLDOS}).
The featureless RG side is thus not a surface projection artifact of the surface-sensitive STM measurement.
The buried modes are not backscattered by the domain wall, in agreement with the small $R(E)$ calculated for $\stk{+^{5}}$ incidence.
Where the STM is important is in coupling to the top layers, similarly to the graphene leads in the WPD calculation.
This coupling realizes the mode selection in both the experiment and the WPD calculation: without it, most of the asymmetric scattering effect is lost.
We explore this by coupling the graphene leads to the bottom graphene layer, which strongly reduces the reflection asymmetry between the two directions of injection (see Supplementary Section~S5.3, Supplementary Fig.~S6 and Supplementary Video~2).

The main effect which leads to the reflection asymmetry is that an incoming wave packet from the $\stk{+^{5}}$ $E \sim k^6$ band can easily couple to the $E \sim k^5$ band of the $\stk{+^{4}-}$ structure, but this mode is mostly invisible to the STM tip (see panel V of Fig.~\ref{fig:wpd}).
From the other direction, an incoming wave packet from the Dirac band couples less effectively to the only $\stk{+^{5}}$ band available.
It is worth exploring this effect from a scattering perspective.
It is important to note that the asymmetric reflection does not conflict with the reciprocity of scattering~\cite{Buttiker1986-bg, Buttiker1988-lz}.
Reciprocity guarantees that the transmission probability between any pair of modes is the same in the two directions.
The STM tip injects and probes through the top graphene layer, and so addresses only the modes with weight there: on the \stk{+^{4}-} side the Dirac band (state VI in Fig.~\ref{fig:wpd}b), while the band marked V, with negligible weight on the top layer, is hidden from the tip; on the \stk{+^{5}} side the flat band and its dispersive part (states I and II in Fig.~\ref{fig:wpd}a).
For such a subset of the modes reciprocity imposes no constraint on the reflections: the reflection seen within it can be strong on one side and absent on the other, as long as the difference is carried by the modes hidden from the probe.
In Supplementary Section~S6 we show this in a simple scattering model of the junction: the difference between the reflections on the two sides is exactly matched by the difference in what leaks into the hidden modes.
In our case the hidden mode is the band marked ``V'' in Fig.~\ref{fig:wpd}. 
% Throughout, the scattering is elastic and time-reversal symmetric: none of the conditions listed in the introduction is broken.

\section*{Conclusions}

We presented STM measurements of stacking domain wall solitons in six-layer graphite, separating a $\stk{+^{5}}$ hexalayer RG from a $\stk{+^{4}-}$ stacking fault region.
Within the surface-projected gap of the RG, quasiparticle interference appears exclusively on the stacking fault side of the walls, where the reflected Dirac band states form standing waves between two walls acting as partially reflecting mirrors.
The RG side remains featureless in the experiment, and our layer resolved tight binding calculations show interference in none of the six graphene layers.
Wave packet dynamics simulations reproduce this directional reflection asymmetry and trace it to the mismatch in the layer polarization of the modes on the two sides, made observable by the top layer coupling of the STM tip.

The mirrors of the cavity are defined by the crystal structure alone: a single-layer shift in stacking, with no lithography and no fabrication disorder.
The mode selection supplied here by the STM tip can be built into devices, since point contacts through holes in an encapsulating BN layer address the top graphene layer directly~\cite{Handschin2015-ya, Telford2018-yd}.
The asymmetry would become observable in charge transport measurements in a three-terminal geometry: top-layer contacts on the two sides of a wall and a third contact, coupled to the bottom layer or to all layers of the $\stk{+^{4}-}$ side, which gives the band hidden from the top layer (state V in Fig.~\ref{fig:wpd}b) a terminal of its own (see discussion in Supplementary Section~S6).
The wall routes the two inputs to different outputs with every pairwise transmission symmetric, as the photonic gratings do with their diffraction orders~\cite{Serebryannikov2009, Foteinopoulou2022}.
Such a junction is a building block for a one-way mirror that is quantum coherent, deterministic, and intrinsic to the crystal, in contrast to ballistic rectifiers defined by etched geometry~\cite{Auton2016-hs}.
Since the domain walls can be displaced under electric fields and mechanical stress~\cite{Yankowitz2014-bf, Jiang2018-ci, Yeo2025-qf}, the mirror itself is in principle movable and switchable, opening new perspectives for the realization of electron optics in rhombohedral graphite.

\subsection*{Methods}

\subsubsection*{Exfoliation and AFM characterization of samples}

We exfoliated the samples using ``blue tape'' (Ultron systems P/N: 1008R-8.0).
As substrate we used Si wafers (90 nm SiO$_2$).
Graphite samples were purchased from NGS Trading \& Consulting GmbH (www.graphit.de).
We performed AFM measurements using an NX10 microscope from Park Systems in tapping mode (non-contact mode).

\subsubsection*{STM measurements}

STM measurements were carried out at a temperature of 9~K, using an instrument from RHK (PanScan Freedom), with a base pressure of 5$\times$10$^{-11}$ Torr.
Samples investigated by STM were exfoliated flakes, supported on a Si/SiO$_2$ substrate and contacted using In spikes~\cite{Girit2007-vi}.
dI/dV maps and line spectra were measured with a tunneling current setpoint of $145$~pA; the tip height was stabilized on the topography at a sample bias of $-0.5$~V during the acquisition of the maps.
Measurements were conducted with a lock-in amplifier at a frequency of 1372~Hz, with a modulation amplitude of 3.75~mV.
Data analysis of the STM measurements was carried out using the open-source Python tool: RHKPY~\cite{Nemes-Incze2023-fd}.

\subsubsection*{Calculation details}

\subsubsection*{Tight binding model}
To determine the electronic properties of the systems we used the same distance dependent tight-binding model as in Ref.~\cite{Balogh2026-ni}. The tight-binding Hamiltonian for electrons in a single sheet of graphene considering that electrons can hop between first, second and third nearest neighbours has the following form:
\begin{equation}
\hat{\mathrm{H}}_{\mathrm{TB},\mathrm{in-plane}}=-\gamma_1\sum_{\langle i,j \rangle}\left(a^{\dagger}_{i}b_{j} + \mathrm{H.c.}\right) -\gamma_2\sum_{\langle\langle i,j \rangle\rangle}\left(a^{\dagger}_{i}a_{j} + b^{\dagger}_{i}b_{j}+ \mathrm{H.c.}\right)-\gamma_3\sum_{\langle\langle\langle i,j \rangle\rangle\rangle,}\left(a^{\dagger}_{i}b_{j} + \mathrm{H.c.}\right)
\label{eq:TB_Ham_inplane}
\end{equation}
The operators $a^{\dagger}_{i}\ (a_{i})$ create (annihilate) an electron on sublattice A, site $i$. The notation is the same for the $b$ operators. $\langle\cdots\rangle$ denotes summing over the first-, $\langle\langle\cdots\rangle\rangle$ denotes summing over the second- and $\langle\langle\langle\cdots\rangle\rangle\rangle$ denotes summing over the third-nearest neighbours. The hoppings have the following form~\cite{gorog}:
\begin{equation}
\gamma_i(r) = \gamma_{i,0}\cdot \mathrm{e}^{\beta_i\left(\frac{r}{a_{\mathrm{CC}}}-1\right)}
\label{eq:inplane_hopping}
\end{equation}
where the subscript $i=1,2,3$ denotes the first, second and third neighbours, $\gamma_{i,0}$ is the value of the hopping integral in the equilibrium position, $\beta_i$ is the strength of the decay of the hopping integral and $r$ is the distance between the two atoms. Both $\gamma_{i,0}$ and $\beta_i$ were obtained from fitting the tight-binding band structure to DFT results using the least-squares method.
To include the hoppings between the different layers, the Hamiltonian~\eqref{eq:TB_Ham_inplane} was extended with the following terms:
\begin{equation}
\hat{\mathrm{H}}_{\mathrm{TB},\mathrm{interlayer}}=\sum_{i,j}\gamma_{ij}c^{\dagger}_{i}c_{j}+\mathrm{H.c.}
\end{equation}
The operators $c^{\dagger}_{i},\ (c_{i})$ create (annihilate) an electron on site $i$. Sites $i$ and $j$ are in adjacent layers. The interlayer hopping between atoms at positions $\mathbf{r}_i$ and $\mathbf{r}_j$ takes the form proposed by Slater and Koster~\cite{Slater_Koster}:
\begin{equation}
\gamma_{ij}=\sum_i\left[\left(\Tilde{V}_{pp\pi}\cdot \cos^2(\theta_i)+\Tilde{V}_{pp\sigma}\cdot \sin^2(\theta_i)\right)\cdot\mathrm{e}^{-\alpha\cdot r_{ij}}\cdot\Theta(\mathbf{r}_{ij})\right]
\end{equation}
The spin indices are omitted in the formulas above. The hopping parameters were fitted to mono- and bilayer graphene band structures calculated using the SIESTA code~\cite{siesta1,siesta2,siesta3}. For details of the fitting procedure see Ref.~\cite{Balogh2026-ni}.

The SiO$_2$ substrate dopes the sample. To take this into account in the TB calculations, on-site terms were added to the Hamiltonian. For details of the modelling of the substrate-induced doping see Supplementary Section~S3.

\subsubsection*{DFT calculations}

Density functional theory (DFT) calculations were performed using the projector augmented wave (PAW) method~\cite{Blochl1994} as implemented in the Vienna ab initio simulation package (VASP)~\cite{Kresse1993}.
The exchange--correlation interactions, including nonlocal van der Waals interactions, were described using the optB86b-vdW functional~\cite{Klimes2011}.
A plane-wave basis set with a kinetic energy cutoff of 500~eV was used.
The six-layer graphene unit cells, in both the \stk{+^{5}} (ABCABC) and the \stk{+^{4}-} (ABCABA) stacking configurations, were modeled using a slab geometry.
A vacuum spacing of 16~\AA\ was applied perpendicular to the graphene plane to prevent interactions between periodic images.
Atomic positions were relaxed using the conjugate gradient algorithm until all forces were smaller than 0.01~eV/\AA.
The Brillouin zone was sampled using a $36 \times 36 \times 1$ Monkhorst--Pack $k$-point grid~\cite{Monkhorst_Pack}.

\subsubsection*{Molecular dynamics calculations}

The model systems were constructed as initially flat six-layer graphene ribbons with \stk{+^{5}} stacking. 
The ribbon was oriented with the longitudinal $y$-axis along the zigzag direction and the transverse $x$-axis along the armchair direction, with periodic boundary conditions applied only along $x$. 
Stacking domain walls were introduced by imposing a spatially varying in-plane displacement on the top graphene layer along an armchair direction forming an angle of $30^{\circ}$ with the $y$-axis. 
The total displacement corresponded to an integer multiple of the carbon-carbon bond length, while its spatial variation was described by a smooth sigmoid-like step profile. 
This construction produces a continuous change in stacking registry and confines the associated strain to a finite domain-wall region. 
The initial domain-wall width was set to 15~nm, comparable to the $\sim$10~nm width observed experimentally. 
Because the displacement direction is not perpendicular to the domain boundary, the resulting solitons contain both shear and tensile components, corresponding to the mixed-type domain walls commonly observed experimentally. 
The resulting structures were subsequently relaxed using molecular statics calculations in LAMMPS~\cite{THOMPSON2022108171} with the conjugate-gradient minimization algorithm. 
Intralayer carbon interactions were described by the Long-range Carbon Bond-Order Potential (LCBOP)~\cite{Los2003-intrinsic}, while registry-dependent interlayer interactions were treated using the Dihedral-angle-corrected Registry-dependent Interlayer Potential (DRIP)~\cite{Wen2018-Dihedral}. 
During minimization, atomic relaxation was restricted to the 15 nm domain-wall regions, whereas the surrounding \stk{+^{5}} and \stk{+^{4}-} domains were kept fixed to prevent their transformation toward the energetically preferred Bernal stacking. 
Relaxation resulted in pronounced out-of-plane corrugation localized around the stacking domain boundaries.

\subsubsection*{Wave packet dynamics calculations}

Wave packet dynamics (WPD) calculations followed the method of Ref.~\cite{Mark2025-ek}.
The time-dependent Schr\"odinger equation was solved using the split-operator Fourier transform method, and a time--energy Fourier transform yielded the energy-resolved wave function.
The Hamiltonian was constructed from a local carbon pseudopotential, a sum of Gaussians centered on the atomic sites, developed for van der Waals stacks of carbon sheets and validated against the band structures of AA, AB, and ABC stackings~\cite{Mark2023-tr}.
The model device consisted of a single-layer graphene lead, the \stk{+^{5}} and \stk{+^{4}-} domains, and the relaxed wall between them; the system was periodic in the direction parallel to the wall, with free-space boundary conditions at the two ends, so that outgoing wave packets were not reflected back into the simulation region.
The initial state was constructed from the Bloch states of the graphene lead by a Gaussian convolution of states around the K point, with momentum perpendicular to the wall: a planar wave front injected into the outermost layer of the junction.
The K valley alone is sufficient: the domain wall is smooth on the atomic scale and does not scatter between the valleys, and time-reversal symmetry maps the K wave packet onto its K$'$ counterpart with the same reflection probabilities.
The total reflection probability was obtained by time-integrating the probability current returning into the lead, and the energy-resolved reflection $R(E) = 1 - I_{\mathrm{out}}(E)/I_{\mathrm{in}}(E)$ was normalized to a reference calculation without the domain wall~\cite{Mark2025-ek}.

\subsection*{Acknowledgements}

	The work was conducted within the framework of the MTA - HUN-REN EK Lendület “Momentum” Topology in Nanomaterials Research project LP2024-17.
	Funding from the National Research, Development, and Innovation Office (NKFIH) in Hungary, through the Grants K-146156, PD-146479, K-134258, FK-142985 and Excellence 151372 is acknowledged.
	P.V., M.S. and A.P. acknowledge support from the J\'{a}nos Bolyai Research Scholarship of the Hungarian Academy of Sciences.
	K.K. was supported by the \'{U}NKP-23-3-II-BME-161 New National Excellence Program of the Ministry for Culture and Innovation from the source of the National Research, Development and Innovation Fund.
	A.M. is funded as Research Associate by the Fund for Scientific Research (FSR-F.N.R.S.) of Belgium, which also supported G.I.M. and P.V.
	M.S. acknowledges the Digital Government Development and Project Management Ltd. for awarding access to the Komondor HPC facility based in Hungary.

\subsection*{Author contributions}
STM measurements were conducted by K.K.
K.K. and P.N.-I. analyzed the STM data.
A.B. performed the tight-binding calculations under supervision from Z.T. and J.K.
P.V. and M.S. performed the DFT calculations.
D.V., K.M. and K.K. prepared the samples; A.P. and K.K. performed the electronic Raman scattering measurements.
M.S. performed the molecular dynamics calculations.
G.I.M. and P.V. performed the wave packet dynamics calculations with help from M.S. and A.M.
L.T. and J.C. aided in data analysis and interpretation.
P.N.-I. conceived and coordinated the project.
P.N.-I. and K.K. wrote the manuscript with input from all authors.

\subsection*{Data availability}

	The raw data and Python scripts used to generate the figures in the main text are available at Zenodo: dx.doi.org/zzzzzz

\subsection*{Competing interests}

	The authors declare no competing interests.

\newpage

\bibliography{confinement}

\end{document}